# Gender gap on concept inventories in physics: What is consistent, what is inconsistent, and what factors influence the gap?

Adrian Madsen,[1] Sarah B. McKagan,[1] and Eleanor C. Sayre[2]

[1]*American Association of Physics Teachers, College Park, Maryland, USA*
[2]*Kansas State University, Manhattan, Kansas, USA*

We review the literature on the gender gap on concept inventories in physics. Across studies of the most commonly used mechanics concept inventories, the Force Concept Inventory and Force and Motion Conceptual Evaluation, men's average pretest scores are always higher than women's, and in most cases men's posttest scores are higher as well. The weighted average gender difference on these tests is 13% for pretest scores, 12% for posttest scores, and 6% for normalized gain. This difference is much smaller than the average difference in normalized gain between traditional lecture and interactive engagement (25%), but it is large enough that it could impact the results of studies comparing the effectiveness of different teaching methods. There is sometimes a gender gap on commonly used electricity and magnetism concept inventories, the Brief Electricity and Magnetism Assessment and Conceptual Survey of Electricity and Magnetism, but it is usually much smaller and sometimes is zero or favors women. The weighted average gender difference on these tests is 3.7% for pretest scores, 8.5% for posttest scores, and 6% for normalized gain. There are far fewer studies of the gender gap on electricity and magnetism concept inventories and much more variation in the existing studies. Based on our analysis of 26 published articles comparing the impact of 30 factors that could potentially influence the gender gap, no single factor is sufficient to explain the gap. Several high-profile studies that have claimed to account for or reduce the gender gap have failed to be replicated in subsequent studies, suggesting that isolated claims of explanations of the gender gap should be interpreted with caution. For example, claims that the gender gap could be eliminated through interactive engagement teaching methods or through a "values affirmation writing exercise" were not supported by subsequent studies. Suggestions that the gender gap might be reduced by changing the wording of "male-oriented" questions or refraining from asking demographic questions before administering the test are not supported by the evidence. Other factors, such as gender differences in background preparation, scores on different kinds of assessment, and splits between how students respond to test questions when answering for themselves or for a "scientist" do contribute to a difference between male and female responses, but the size of these differences is smaller than the size of the overall gender gap, suggesting that the gender gap is most likely due to the combination of many small factors rather than any one factor that can easily be modified.

## I. INTRODUCTION

Concept inventories are research-based multiple-choice assessment instruments designed to test students' conceptual understanding of a topic. Physics instructors often use concept inventories to gauge their students' understanding of physics concepts and in turn the effectiveness of their instruction [1,2]. Studies showing dramatic differences in concept inventory scores between traditional lecture classes and interactive engagement [3] have had a major impact on physics education reform by convincing many instructors to change their teaching methods [4]. The most commonly used concept inventories in physics are the Force Concept Inventory (FCI) [5] and the Force and Motion Conceptual Evaluation (FMCE) [6] for introductory mechanics and the Brief Electricity and Magnetism Assessment (BEMA) and the Conceptual Survey of Electricity and Magnetism (CSEM) [7] for introductory electricity and magnetism (E&M). Male students almost always outperform female students on these types of standardized conceptual multiple-choice assessments. We call this difference in scores the "gender gap." Across previously published data, the weighted average gender gap for the two different mechanics concept inventories is 13% for the pretest, 12% for the posttest, and 6% for the normalized gain. There is more variability in the size of the gap across different institutions, instructors, teaching methods, etc. on the posttest and in the normalized gain than on the pretest. The weighted average gender gap for the two different electricity and magnetism concept inventories is 3.7% for the pretest, 8.5% for the posttest, and 6% for the normalized gain. Compared to the mechanics concept inventories, the electricity and magnetism concept inventories have more variation in both the pretest and the posttest across fewer studies; therefore, it is more difficult to see patterns in these studies.

The existence of a gender gap on the pre- and posttest of these concept inventories brings up many important questions. For example, is the gender gap an artifact of the testing format or is it due to a real difference between the genders' understanding of the concepts that the tests are designed to measure? Faculty members are likely to ask, "is the gender gap for my students comparable to the gap elsewhere?" Researchers likely have questions of a different nature: for example, "does the gender gap have a substantial impact on the results of my study?"

Numerous studies have investigated the gender gap on these concept inventories and other measures in physics. These studies have looked at how various factors influence the gender gap and whether different techniques can reduce the gap. In this paper, we present a synthesis of the research done on the gender gap using concept inventories. We address questions relevant to instructors giving concept inventories in their courses and researchers using these assessments in their studies. We start with an overview of the pretest, posttest, and normalized gain gender gaps and then discuss the numerous factors that have been investigated to influence the gender gap and the direction and strength of the influence. We conclude with important takeaways for instructors and researchers and a discussion of open questions.

We find that the story of the gender gap in physics is not a simple and clean-cut one. Many physics educators and researchers are familiar with the story of how interactive teaching methods were enacted in the late 1980s and scores on concept inventories increased drastically as compared to traditional instructional methods. The positive effect of interactive engagement teaching methods has continually been found to be substantial and relatively consistent across institutions, instructors, and students. In contrast, the story of what influences the gender gap and how to reduce it is much less clear.

## II. PRETEST GAPS

Studies that compare male and female students' scores on the FCI and FMCE contain a remarkably consistent finding: there is always a gender gap favoring men on the pretest [8–19]. This gap ranges between 8.2% and 18.7% (see Fig. 1), with a weighted average value of 13.0%. This gender gap occurs at different institutions with different instructors and student populations across a wide range of pretest scores. Additionally, women, on average, do worse on every question on the FCI, though how much worse varies by question [11,20].

The weighted average pretest gender gap favoring men on the BEMA and the CSEM is 3.7%, which ranges from −0.2% to 7.1% [16,21–23], much smaller than that on the FCI and FMCE. In some cases there is no gender gap at all. (The weighted pretest gender-gap value of 3.7% includes the five studies on the BEMA and CSEM pictured in Fig. 1

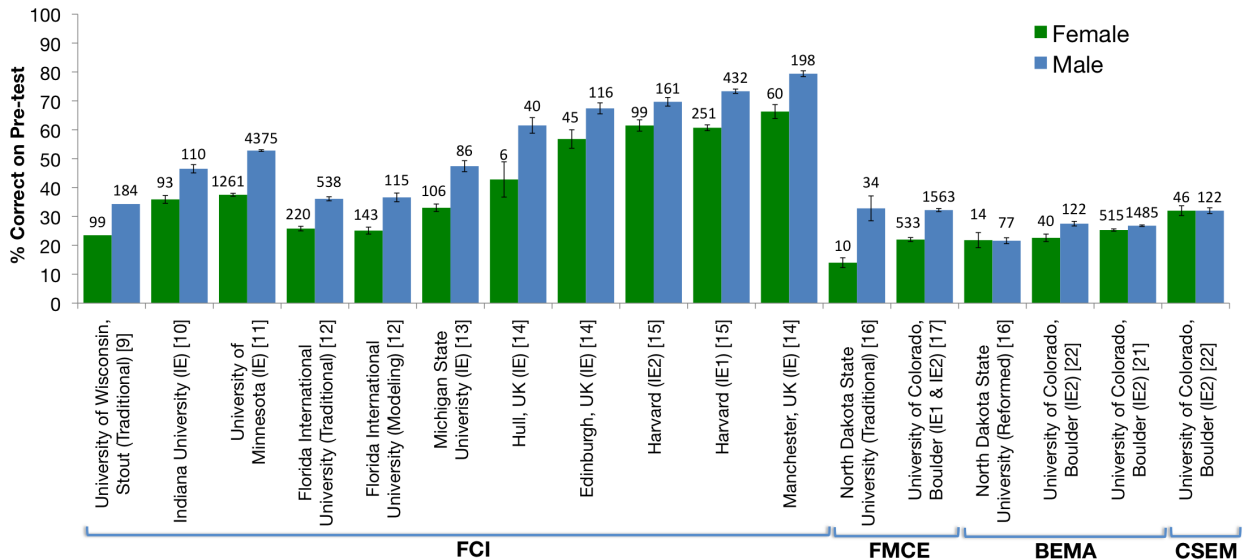

FIG. 1 (color online). Pretest scores for men and women across several institutions on the FCI, FMCE, BEMA, and CSEM. Pretest gender gap on the FCI is usually about 10% across intuitions, instructors, countries, etc. Numbers above bars indicate number of men and women included in the study at each institution. Error bars represent the standard error. If error bars are not present, they were not reported in the study. Instructional methods used in each course are given in parentheses (although these should not influence pretest scores, they are included here to make comparisons with subsequent figures easier). IE is interactive engagement. IE1 and IE2 are different levels of interactive engagement, defined as partially interactive and fully interactive, respectively [17]. This graph does not include CSEM data from Ref. [28], which reported only gender gaps and not scores.

as well as CSEM data from studio and nonstudio courses studies by Kohl and Kuo [23]. These data are not included in Fig. 1 because gender gaps were reported without average pretest scores.) The pattern in test scores for E&M tests is different from mechanics tests. The pretest scores in mechanics represent substantial prior intuitive knowledge of the topics as evidenced by a widespread distribution of scores. The pretest scores for E&M tests are likely subject to a ''floor effect''; i.e., the test cannot discriminate between groups because all students score poorly due to their lack of familiarity and experience with these topics [21]. This idea is supported by consistently low pretest scores on E&M concept inventories with a lack of spread in score distribution [22,24]. It may not be as meaningful to look at the pretest gender gap on E&M tests as compared to mechanics tests because neither gender comes into the course with adequate knowledge about electricity and magnetism for the test to measure.

## III. POSTTEST GAPS

There is usually a gender gap on the posttest for the FCI [9–15], FMCE [16,17,25], BEMA [16,21,22], and CSEM [22,23] though the size of the gender gap is much more variable than that on the pretest (Fig. 2). In most cases there is still a gender gap on the posttest, but the size of the gap varies. On the FCI and FMCE the weighted average posttest gender gap is 11.6% and ranges from 1.5% to 24.6% [9–15]. The weighted average posttest gender gap on the BEMA and CSEM is 8.5% with a range from −3.3% to 13%. (The weighted average includes data from the three studies on the BEMA and CSEM pictured in Fig. 2 as well as CSEM data from studio and nonstudio courses studies by Kohl and Kuo [23]. These data are not included in Fig. 2 because only gender gaps were reported, not average posttest scores.)

We can also compare the pretest and posttest gender gaps to look at how the gap changes over the course of the semester (Fig. 3). We would hope that the gender gap decreases from pre- to posttest, or at minimum stays the same, but we find that the way the gender gap changes varies greatly across studies. Several studies have found that the FCI gender gap increases from pre- to posttest [9,10,12] with the increase ranging from 1.2% to 8.7%. Other studies have found the FCI gender gap decreases from pre- to posttest [11,13–15] with the decrease ranging from 0.6% to 8.6%. There are two studies that report on the FMCE gender gap. In one study the gender gap increased by 5.8% from pre- to posttest [16]. In the other there was no change in the FMCE gender gap from pre- to posttest when the scores were averaged over many semesters although the gender gap increased and decreased from pre- to posttest in single semesters [17]. This difference in the change in gender gap from pre- to posttest over single semesters may be related to differences in instructors [17] although it has been noted that these types of differences are consistent with statistical fluctuations [11]. On the BEMA the gender gap increases from pre- to posttest with the increase ranging from 4.8% to 8.6% [21,23] while the gender gap on the CSEM both increases and decreases from pre- to posttest

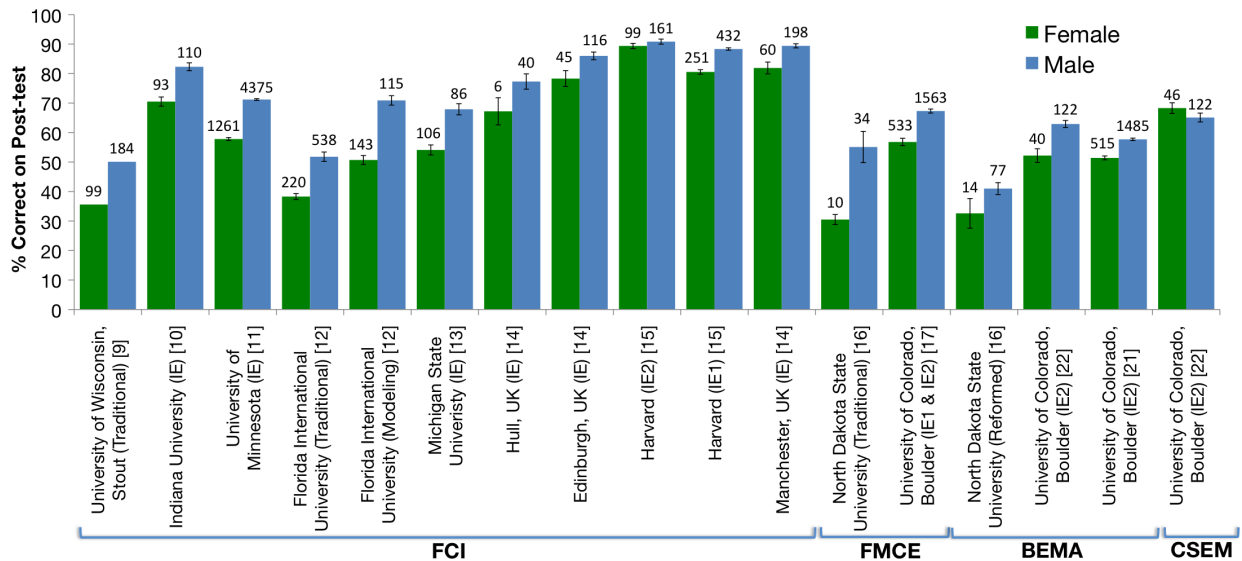

FIG. 2 (color online). Posttest scores for men and women across several institutions on the FCI, FMCE, BEMA, and CSEM. Posttest gender gap is more variable than pretest gender gap. Numbers above bars indicate number of men and women included in the study at each institution. Error bars represent the standard error. If no error bars are present, they were not reported in the study. Instructional methods used in each course are given in parentheses. This graph does not include CSEM data from Ref. [28], which reported only gender gaps and not scores.

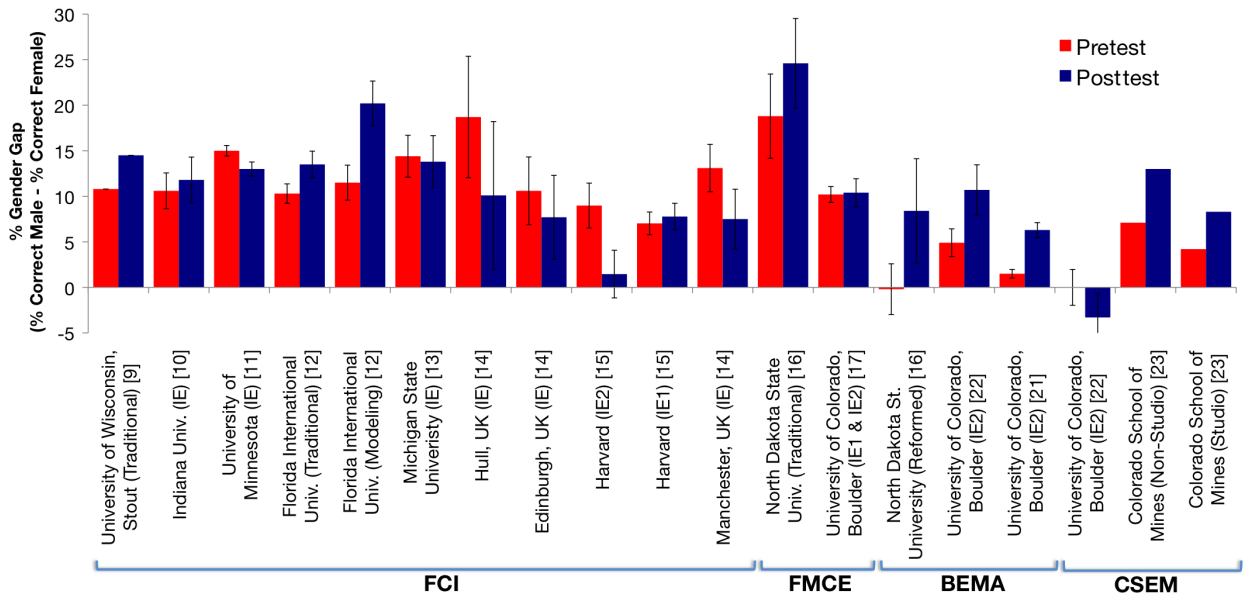

FIG. 3 (color online). Gender gap in percentage on pre- and posttest across several institutions on the FCI, FMCE, BEMA, and CSEM. The gender gap changes differently in each situation. Sometimes it increases from pre- to posttest, other times it decreases or stays the same. The way the gender gap changes from pre- to posttest does not seem to be related to type of instruction (interactive engagement or traditional). Error bars represent the standard error. If no error bars are present, they were not reported in the study.

with a range from −3.3% to 5.9% [22,23,26]. Once again, the changes in gender gap from pre- to posttest on the CSEM and BEMA may not be meaningful because of the floor effect on pretest scores. In summary, the posttest gender gaps and the change in gender gap from pre- to posttest show a large amount of variation across studies.

## IV. NORMALIZED GAIN GAPS

Normalized gain is a conventional metric used to compare the effectiveness of educational interventions [3]. It is important to look at the gender gap in normalized gain in order to determine how such a gap might affect these comparisons (see Fig. 4). The magnitude of the normalized

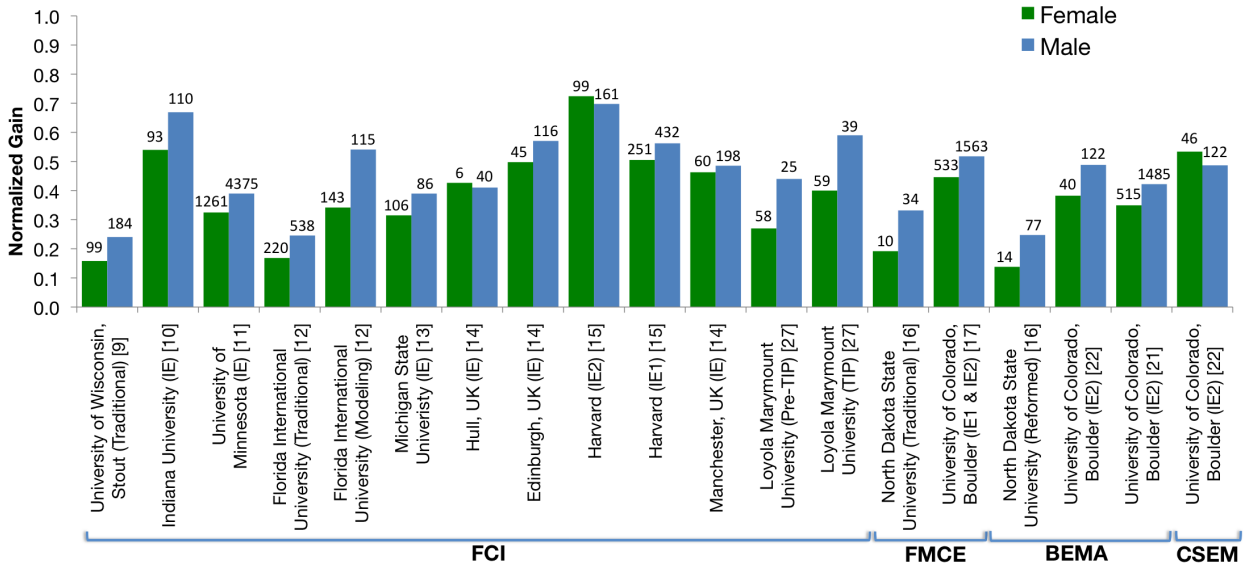

FIG. 4 (color online). Gender gap in normalized gain on the FCI, FMCE, BEMA, and CSEM across several institutions. The magnitude of male students' normalized gain is usually greater than female students'. The average difference in normalized gain between men and women is 0.06, much smaller than the average difference in normalized gain between interactive engagement and traditional teaching methods [3]. TIP is Thinking in Physics pedagogy [27] and Pre-TIP refers to classes taught by the same instructor before the TIP teaching methods were implemented. Numbers above bars indicate the number of men and women included in the study at each institution.

gain is usually greater for men than for women, with the normalized gain gender gap across institutions showing the same kind of variation as the posttest gender gap. The weighted normalized average gain for mechanics concept inventories is 0.43 for men and 0.37 for women (these averages include data from Coletta *et al.* [27] as normalized gain was reported in this paper, but not pre- or posttest scores). The weighted average difference is 0.06 with a range from −0.05 to 0.20. For E&M concept inventories, the weighted normalized average gain is very similar and has a value of 0.42 for men and 0.36 for women. The average weighted difference is 0.06 (these averages do not include the CSEM data from Kohl and Kuo [23]). If we look back to the story of interactive engagement versus traditional teaching methods, Hake [3] found an average normalized gain of $0.23 \pm 0.04$ standard deviation (SD) for 14 "traditional" courses and $0.48 \pm 0.14$ SD for 48 "interactive engagement" courses, making the difference in normalized gain by teaching method 0.25. The difference in normalized average gain between male and female students is much smaller than the difference in normalized average gains between traditional and interactive engagement teaching methods, but large enough that it could have an impact on studies using normalized gain to compare the effectiveness of different teaching methods.

In summary, the gender gap on the posttest, the way the gender gap changes from pre- to posttest, and the gender gap on normalized gain all show significant variation across studies. The gender gap on normalized gain is much smaller than the difference between interactive engagement teaching methods and traditional teaching methods, implying that teaching methods have a substantially greater impact on concept inventory scores than gender differences.

## V. WHICH FACTORS MAY CONTRIBUTE TO THE GENDER GAP AND WHAT IS KNOWN ABOUT HOW THEY INFLUENCE THE GENDER GAP?

With a clearer understanding of the findings on the pretest, posttest, and normalized gain gender gaps on concept inventories, we next ask, "what factors influence this gender gap and in what way?" There are many factors that have been postulated to influence the gender gap in physics concept inventories, including bias in the test questions themselves, background and preparation of students, and stereotype threat. Table I shows a list of the factors that have been investigated, their demonstrated impact, and the concept inventories in which they have been tested. Some of these factors have shown a consistent effect on the gender gap across studies, some have consistently been shown not to have an effect on the gender gap, others have shown inconsistent results, and there are some factors which have not been investigated thoroughly enough to allow conclusions to be drawn about their effect. Several high-profile studies that have claimed to account for or reduce the gender gap have failed to be replicated in subsequent studies, suggesting that isolated claims of explanations of the gender gap should be interpreted with caution. Below we describe these high-profile studies and the factors that do and do not influence the gender gap on concept inventories in physics. Though a large number of factors have been investigated to explain the gender gap, no one factor has been able to fully account for the gap. This leads us to believe that the gender gap is a complex phenomenon that cannot be easily explained. It is likely that many of these factors contribute to the gender gap, but not in a way that is easy to observe and quantify.

### A. Background and preparation

One hypothesis is that differential background and preparation between men and women is a major contributor to the gender gap on these concept inventories in physics. Perhaps men have stronger preparation and thus outperform women on the pretest. Then, because men start the physics course with more knowledge, they are able to gain more from the physics course (in some cases), or in other cases, men and women learn the same amount over the course of the semester, but men start out with an advantage that persists through the course.

Kost *et al.* [17,21,28] used a regression model to explore how background factors accounted for the gender gap in the FMCE and BEMA posttest scores. These factors are listed in Table I as Refs. [17,21,28], respectively. Their model predicted how much of the gender gap in posttest scores could be accounted for by factors other than gender. They found several factors that accounted for a substantial portion of the gender gap. For posttest FMCE scores [17], the final regression model accounted for 7.5 points of the 10.7 point gender gap or 70% of the gap. Predictive variables included FMCE pretest, combined math score (ACT, SAT and/or math placement exam), Colorado Learning Attitudes about Science Survey (CLASS) pretest, and semester the course was taken. Noncontributing factors included years of high school physics, years of high school calculus, high school grade point average (GPA), declared major, and ethnicity.

Using a similar regression model for the BEMA posttest scores, they were able to account for 4.2 points of the 6.8 point gender gap, or 62% of the gap. Predictive variables in this model included BEMA pretest, FMCE posttest (from Physics 1), combined math score, CLASS pretest, and semester the course was taken. Similar to the FMCE results, they found that years of high school physics, declared major, and ethnicity were not predictive variables for the BEMA posttest. Kost *et al.* [28] carried out an additional regression analysis to determine if physics self-efficacy and identity could predict FMCE posttest scores and course grades. They found that favorable ratings on both self-efficacy and identity survey questions were useful predictors for grades, but these variables did not predict FMCE posttest scores beyond factors already included in their regression model. In both of these studies, most of the gender gap in posttest scores can be accounted for with a handful of background variables.

TABLE I. List of factors investigated that may influence the gender gap in pre- and/or posttest scores and the associated concept inventory studied.

| Type of factor | Factor investigated | Demonstrated impact | | | Test |
|---|---|---|---|---|---|
| | | Yes | No | Inconclusive | |
| Background and preparation | High school GPA | | x[a] | | BEMA [21], FMCE [20] |
| | Years high school physics | | x[a] | | BEMA [15], FMCE [17] |
| | Declared major | | x[a] | | BEMA [15], FMCE [17] |
| | Years high school calculus | | x[a] | | FMCE [17] |
| | Physics 1 GPA | | x[a] | | BEMA [21] |
| Other assessment | Lawson test of scientific reasoning | | | x | FCI [27] |
| | Problem-solving pretest | x[b] | | | FCI [18] |
| | Free-response conceptual pretest | x[b] | | | FCI [18] |
| | CSEM pretest | | x | | CSEM [23] |
| | University diagnostic placement math exam | x | | | FMCE [17] |
| | FMCE pretest | x | | | FMCE [17,28] |
| | FMCE posttest (from Physics 1) | x | | | FMCE [17] |
| | BEMA Pretest | x | | | BEMA [21] |
| | Physics 1 average exam score | x | | | |
| | ACT math score | x | | | FMCE [17,29], BEMA [21] |
| | SAT math score | x | | | FMCE [17,29], BEMA [21], FCI [15] |
| Teaching method or instructor | Thinking in Physics pedagogy | | x | | FCI [27] |
| | Level of interactive engagement | | | x | CSEM [26], FCI [15], FMCE [17,25] |
| | Studio physics | | x | | CSEM [23] |
| | Instructor | | | x | CSEM [26], BEMA [21], FMCE [17], FCI [11] |
| | Modeling Instruction | x | | | FCI [12] |
| Sociocultural factors | Self-affirmation writing exercises | | | x | FMCE [29,30] |
| | Level of endorsement of gender stereotype | x | | | FMCE [29] |
| | Self-efficacy | | x[a] | | FMCE [28] |
| | Identity | | x[a] | | |
| | CLASS pre- and posttest | x | | | FMCE [17], BEMA [21] |
| | Students' rating of belief in their answer | x | | | FCI [31] |
| | Locus of control over their own grades | x[b] | | | FCI [18] |
| Question construction | Item analysis by student ability (IRT and Rasch methods) | | x | | FCI [20,32] |
| | Everyday and feminine question contexts | | x | | FCI [8,9] |

[a]Not more predictive than other factors already included in regression model. See reference for more details.
[b]Together with seven other measures included in the analysis. See reference for more details.

Antimirova et al. [33] also investigated how background factors influenced students' pre- and posttest scores on the FCI using regression analysis. They reported factors that predicted FCI posttest scores, rather than factors that accounted for the gender gap in scores as Kost et al. [17,21] did in their studies. They found that background factors accounted for more of the variation in scores on the pretest than the posttest. Variables that significantly contributed to the model for the pretest were having taken a high school physics course and being born in Canada (where the university was located). For the posttest, being born in Canada and having additional education beyond high school were found to significantly contribute to the model. Notably, they found that gender did not predict FCI posttest score. They also found that visible minority status, completing high school physics, age, parents' university education status, and non-English home language did not predict FCI posttest scores. These findings indicated that background variables other than gender lead to men outscoring women on the FCI posttest, though the models accounted for very little of the variation in posttest scores, so these variables did not contribute to a model very predictive of the data.

Another way to look at how the initial state of the students' knowledge influences their posttest scores is to bin men and women by pretest score and then compare posttest scores of men and women who started the course with similar pretest scores. Kost et al. [17] binned FMCE posttest scores by FMCE pretest scores and found no significant differences in posttest scores between men and women for any bin, though men did have higher raw scores in each bin. There were significantly more women in the lower pretest score bins and more men in the higher pretest score bins. They also found that FMCE pretest score correlated with FMCE posttest score. They concluded that women's low pretest scores combined with the fact that pre- and posttest scores are correlated (or somewhat correlated) was a "dominant source" of the gender gap, consistent with their findings using the regression model analysis.

Kost et al. also binned BEMA posttest scores for those in a second-semester introductory course by FMCE posttest from the associated first-semester introductory course and similarly found no significant differences between the scores of men and women in each bin; men have higher raw scores than women in four of five bins [21]. Here they concluded that FMCE pretest score accounted for a large portion of the gender gap in BEMA posttest scores, again consistent with their findings using the regression analysis. Kohl and Kuo [23] binned students' normalized gain (instead of posttest score) on the CSEM by their CSEM pretest score and found that significant differences between men and women still existed in most bins, with men outscoring women. In this case, men and women with similar pretest did not have similar normalized gains.

Coletta et al. [27] binned FCI normalized gain by Lawson test of scientific reasoning ability scores for two different instructors. They compared the FCI normalized gains for men and women in each of four Lawson score bins. For one instructor, in the two highest Lawson score bins, men statistically significantly outscored women. For the other instructor, there were no statistically significant gender gaps in any bins. The results of binning the FCI normalized gains by scientific reasoning ability were unclear. Scientific reasoning ability may be a factor contributing to the gender gap, but further research is needed.

Another method to look at the influence of preparation on concept inventory scores is to use some measure of preparation as a covariate when comparing the posttest scores of men and women. Brewe et al. [12] used the SAT math score as a covariate when comparing the FCI posttest scores of men and women and found that men still significantly outperformed women when this covariate was included.

Yet another method to investigate the way background and preparation variables influence concept inventory scores is to match pairs of men and women on as many measures of background and preparation as possible and compare their concept inventory scores. Blue matched 20 men and women in a calculus-based introductory physics course based on eight measures: three pretest scores, three high school background characteristics, their year in college, and their locus of control over their grades [18]. There were no statistically significant differences in the FCI posttest scores for the matched pairs of men and women. This suggests that men and women with equivalent background and preparation score similarly on mechanics concepts inventories, though this study included only 20 pairs of students at one institution.

CLASS pretest scores account for some portion of the gender gap on the FMCE and BEMA posttests using regression models [20,27]. Additional work has been done comparing the way attitudes about physics shift from pretest to posttest based on gender using the CLASS. Adams et al. [34] found that in two semesters of a first-semester algebra-based class women had more negative shifts than men when rating agreement with statements in the real-world connections, personal interest, problem-solving confidence, and problem-solving sophistication categories. Women had slightly more expertlike shifts in the sense-making/effort category. Kost et al. [17,21] also looked at how CLASS scores differed by gender. They compared CLASS scores for students in six first-semester introductory calculus-based courses and found that men and women both made negative shifts from pre- to posttest for every category of question, but women made significantly more negative shifts than men in the three problem-solving categories and two conceptual categories. Kost et al. [21] then compared CLASS scores based on gender for students in 10 semesters of a second-semester introductory calculus-based physics course. They found that women had less expertlike pretest scores than men on all categories except sense making, but the shifts in attitudes between pre- and posttest were statistically similar between women and men on all categories except personal interest. In the first semester women made more negative shifts in attitudes on most categories, while in the second semester men and women had similarly negative shifts. Overall, these studies present a consistent result that female students are less expertlike in their attitudes and beliefs about physics than male students and exhibit more negative shifts in beliefs from pre- to posttest. The gender gap on CLASS scores favoring men is consistent with the fact that CLASS scores were found to account for some part of the gender gap on the FMCE and BEMA posttests.

Background variables can account for a substantial portion of the gender gap (70% of the FMCE posttest gender gap and 62% of BEMA posttest gender gap). For the FMCE and BEMA posttest, variables that account for the gender gap include the FMCE or BEMA pretest, combined math score (ACT, SAT, and/or math placement exam), CLASS pretest, and semester the class was taken. Factors that did not account for the gender gap on the FMCE and BEMA posttest include high school physics, ethnicity, declared major, high school calculus, high school GPA, physics self-efficacy, and identity. It is unclear how scientific reasoning ability contributes to the gender gap. The gender gap was not completely closed when SAT scores were used as a covariate in the analysis, though we cannot tell from this analysis if the gender gap was partially explained by the SAT math score or not. The gender gap was completely closed when 20 pairs of men and women were matched on various background variables, though we do not know if this result is unique to this institution. These studies lead us to conclude that the gender gap cannot fully be explained by the background and preparation factors that have been studied, but that differences in preparation and background are major contributors to the gender gap. It is important to note that most of these background factors are other tests, in which students are subject to the effects of test anxiety and stereotype threat, which are known to adversely affect women. We cannot determine if the differences in the background and preparation variables (which explain a large portion of the gender gap) are true differences in preparation or merely artifacts of the testing situation.

### B. Gender gaps on other measures

There is research showing that women do more poorly than men on many kinds of tests including the SAT I Math, GMAT, all sections of the GRE, and several AP tests [35]. Thus, one possible explanation of the gender gap on concept inventories is that it is merely another example of a more general phenomenon. However, our review of the literature suggests that the gender gap on concept inventories is much larger than typical gender gaps on other tests; so, the general phenomenon of women's poor performance on tests in general is insufficient to explain the gender gap on concept inventories.

We can compare the gender gap on in-class exams to that on concept inventories to test this idea. Kost et al. found that over seven semesters men outperformed women on exams by 4.5% [17]. Docktor et al. found that men's final exam scores for 15 semesters were higher by 3.9% [11] while Bates et al. found no statistically significant difference in final exam scores between men and women at three universities for a single semester [14]. Coletta et al. also found no gender differences in exam scores over several semesters [27]. These differences in exam scores between men and women are comparable to the 3.7% gender gap on the BEMA and CSEM pretests, for which it is generally believed that no students have sufficient background knowledge to understand any of the questions. However, they are much smaller than the average gender gaps for physics concept inventories where students are assumed to have some relevant knowledge: 12% for the FMCE and FCI pretest, 13% for the FMCE and FCI posttest, and 8.5% for the BEMA and CSEM posttest.

We can also compare the difference in exam scores to other measures in a physics course. Several studies have reported on the final grades of men and women in introductory physics classes. When looking at grades for one class during a single semester, researchers found no statistically significant gender gap on final grades [11,16,17,27]. When data from several semesters were aggregated, there was a statistically significant gap in final grades favoring men which ranged from 1.5% to 2.8% [11,17], which is smaller than the gender gap in pretest concept inventory scores. Studies have also looked at gender distributions for different components of the class grade. Kost et al. found that women outperform men on homework and participation by 4.5%.

The drop, fail, or withdraw (DFW) rates of men and women have also been compared. Kohl et al. found that the DFW rates of men and women are similar [23]. Brewe et al. compared the rates of success of men and women, which was the ratio of students who earned a C+ or better to those who earned a D+ or lower or who dropped or withdrew [12]. They found men and women had similar rates of success in their physics classes.

To summarize, the gender gap on exam grades (0%–4.5%) is much smaller than the gender gap on physics concept inventories where students are expected to have some background knowledge (8.5%–13%), leading us to believe that taking tests in general or test anxiety may be a factor contributing to the gender gap on concept inventories in physics. Additionally, women outperform men on homework and participation (by 4.5%), men have equal or higher final grades (0%–2.8%), and men and women have similar DFW rates and rates of success.

### C. Difference in personal beliefs and answer "scientist" would give

Studies have found that students have "splits" between their own personal beliefs about the answers to physics questions and the answers they believe a scientist would give, and that these splits are larger for women than for men. One possible explanation for the gender gap on concept inventories could be that women are more likely to answer concept inventory questions based on their personal beliefs about physics and men are more likely to answer based on the way they think a scientist would. However, in one study that looked at gender differences in splits on a concept inventory, the gender difference in splits was much smaller than the gender difference in scores, suggesting that, while this may be a contributing factor, it is insufficient to explain the gender gap.

The difference between men's and women's personal beliefs about physics and what they think a physicist would believe have been studied using two different tests. McCaskey *et al.* [19] gave the FCI as a pretest. Students were asked to indicate the answer they really believed as well as the answer they thought a scientist would give. Women had a higher incidence of "splits" where their personal answer and the answer they believed a scientist would give differed (average number female splits is 8.1, average number male splits is 4.6). Although women more often indicated different answers for personal belief and scientists, the average differences in their personal belief score and scientist score was about 3%, much smaller than the mechanics concept inventory gender gap of 12%–13%. The number of students participating in this study was small, so it should be replicated before drawing concrete conclusions.

Adams *et al.* [34] observed a similar effect when asking students to answer each question on the CLASS posttest twice, once indicating what they thought and again indicating what a physicist would say. They found a bigger difference in favorable responses between women's personal beliefs about physics and what they believe a physicist would say than the difference between men's personal beliefs and what they believe a physicist would say (about 40% for women and 25% for men). This is further evidence that the difference between personal beliefs and the answer they believe a scientist would give is larger for women than for men.

The difference between personal beliefs about physics and students' identification of the scientist answer is bigger for women than men on pre- and posttest versions of two different tests. This difference likely accounts for a small amount of the gender gap on concept inventories, but not the entire gap.

### D. Teaching method

We can also consider the possibility that the use of certain teaching methods influences the gender gap. Specifically, it has been suggested that women receive significant benefit from active learning environments where they are given the opportunity to express their ideas in discussions [15]. If this were so, the gender gap would decrease with increasingly interactive teaching methods [15]. In one high-profile study by Lorenzo *et al.* [15] at Harvard University, as the level of interactive engagement increased, the gender gap on the FCI decreased on the posttest. In the fully interactive class, the FCI posttest gender gap was no longer statistically significant. This was a very promising finding and efforts were made to replicate it. Pollock *et al.* [25] at the University of Colorado compared the FMCE pre- and posttest gender gaps for three semesters of partially interactive and three semesters of fully interactive courses. They found no statistically significant differences between pre- and posttest gender gaps based on level of interactivity. The students in the Lorenzo study had substantially higher pre- and posttest scores than those studied by Pollock *et al.* Docktor *et al.* [11] suggest that high scoring students face ceiling effects, meaning men and women both cannot score any higher, which effectively makes it seem as though the gender gap decreased.

Kohl and Kuo [23] at the Colorado School of Mines compared CSEM pre- and posttest gender gaps for courses taught in a partially interactive manner and those taught using the studio method, which they describe as fully interactive. They found that the pre- and posttest gender gaps were smaller in the studio classes. This is consistent with the findings of Lorenzo *et al.* Pre- and posttest scores were not reported in the Kohl and Kuo study (only differences in scores), so we cannot determine if their results are also subject to ceiling effects. Pollock [22] compared the pre- and posttest gender gaps for students in a second-semester introductory course at the University of Colorado that was taught using fully interactive engagement teaching techniques. In this course, half of the students were randomly assigned to take the CSEM and the other half took the BEMA. There was no statistically significant pre- or posttest gender gap on the CSEM. Pollock did find a statistically significant difference in the pre- and posttest gender gaps on the BEMA, with the gender gap increasing from pre- to posttest. Here, students in the same course taught by the same instructor using fully interactive teaching techniques showed different gender-gap patterns. This finding does not support the claim that the level of interactivity of a course is related to a decrease in the gender gap, as different students in the same course showed no change or increases in the gender gap from pre- to posttest on different tests.

Brewe *et al.* [12] at Florida International compared the FCI pre- and posttest scores for students enrolled in a traditional introductory course and those enrolled in a course taught using Modeling Instruction, an interactive engagement teaching method. The posttest scores for all students were higher in the Modeling Instruction course, but the posttest gender gap was larger in this course as compared to courses taught with traditional teaching methods. This finding is exactly opposite to Lorenzo *et al.* Coletta *et al.* [27] compared the gender gap in normalized gain (instead of pre- or posttest gender gaps) on the FCI before and after implementing the Thinking in Physics pedagogy, an interactive teaching method aimed at helping students develop basic reasoning skills. They found that both men and women had higher normalized gains when the Thinking in Physics teaching methods were used, but the size of the gender gap in normalized gains remained the same. This does not support the findings of Lorenzo *et al.*

In summary, we find that courses taught using interactive engagement techniques exhibit both increases and decreases in the gender gap from pre- to posttest (Fig. 3). So, it is not always true that interactive engagement teaching techniques are related to decreases in the gender gap over the course of a semester. Using these techniques is beneficial to both genders in all cases, but it is unclear whether this benefit is greater for either gender or which details of implementation are likely to have a differential impact.

### E. Stereotype threat

Stereotype threat is "is a concern or anxiety that one's performance or actions can be seen through the lens of a negative stereotype—a concern that disrupts and undermines performance in negatively stereotyped domains" [35]. For example, African American students who were asked to report their race before taking the GRE performed lower than those who were not [36]. Similarly, female students who were asked to mark their gender before taking the AP calculus test performed significantly lower than those who were not [37]. High-math-ability White males underperformed on a math test when the underperformance stereotype of White males as compared to Asian Americans was mentioned briefly, even though no stereotype of low math ability exists for this group [38]. Women drivers who were reminded about the stereotype of women as poor drivers were more likely to hit jaywalking pedestrians in a driving simulator [39]. Stereotype threat has also been studied in physics with mixed results.

Similar to studies on the GRE and AP calculus test, we can investigate if asking students to indicate their gender before taking an assessment has an adverse effect on female students' scores, presumably by some mechanism like stereotype threat. The first three sets of bars in Figs. 1–4 are studies in which gender data were collected from the students while taking the test. In all other studies included in these figures, the researcher accessed the gender data from other sources such as a university database system. When we compare the pretest and posttest gender gap (Fig. 3) between the studies that asked students to indicate their gender and those who did not, we find that there is no definitive pattern: the pre- and posttest gender gap can be large or small independent of how the demographics were collected. We conclude that if the collection of a student's gender information influences their experience of stereotype threat, simply not asking for their gender is insufficient to mitigate this effect, as judged by the pre- and posttest gender gaps.

It could also be that the fraction of women taking a physics class is related to the strength of the stereotype threat. For example, in classes with high proportions of female students women may feel like they belong in the physics class, thus somewhat mitigating stereotype threat. We looked at the FCI pre- and posttest gender gap based on the percentage of women who were reportedly taking the concept inventory (Fig. 5). We used a linear regression model and found no statistically significant relationship between the percentage of women taking the test and the

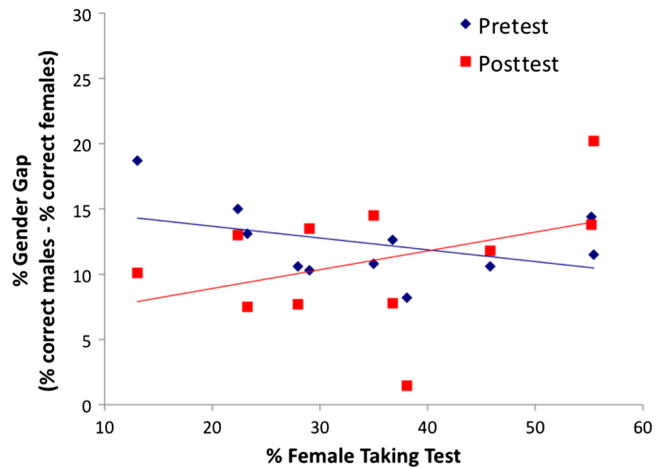

FIG. 5 (color online). Scatter plot summarizing the relationship between the percentage of gender gap on the FCI and the percentage of women taking the pre- and posttest. Each dot represents one of the 10 institutions (though in some cases one institution is represented by two dots, each using a different teaching method). We find no strong relationship between the percentage of women taking the concept inventory and the gender gap in score on either the pre- or posttest.

percent gender gap on the pretest ($R^2 = 0.18$, $p = 0.20$) or the posttest ($R^2 = 0.16$, $p = 0.23$). Thus, there is no indication that a larger fraction of women in a class is related to a smaller gender gap in test scores and a weaker stereotype threat against women. However, we do not have data for any classes with more than 55% women. It is possible that there could be a large effect in classes where a substantial majority of the students are women, for example, at a women's college.

We do not find evidence for stereotype threat using the simple measures discussed above, but it could be that an intervention designed to get at the underlying issue could decrease the negative results of this effect. In a high-profile study at the University of Colorado by Miyake *et al.* [29], students completed a short values affirmation writing exercise twice at the beginning of the semester in a calculus-based introductory physics course. This intervention was found to eliminate the gender gap on the FMCE, decrease the gender gap on exam scores, and increase female students' grades from the C range to the B range, presumably by mitigating the effects of stereotype threat by "reestablish(ing) a perception of personal integrity and worth, which in turn can provide them with the internal resources needed for coping effectively" [29]. The values affirmation exercise was repeated in subsequent semesters where they observed a decrease in the gender gap on exam scores and improvement in women's grades, but they did not find that the gender gap was eliminated for the FMCE [30]. Instead, they found that women in the control condition outperformed the men in the control condition and the women in the value affirmation condition. This result suggests that the values affirmation exercise is not beneficial to female students' FMCE scores, exactly opposite the result found in the original study. A similar study was conducted at North Dakota State University in both semesters of their introductory calculus-based sequence [16]. There were a small number of women in each course, so comparisons could not be made by gender. Instead, the researchers compared grades and FMCE and BEMA scores for groups containing both men and women who had and had not completed the self-affirmation writing exercise. They found that those who had completed the values affirmation writing exercise in the second-semester course had higher normalized learning gains on the BEMA, but this was not found in the first-semester course, where the FMCE was given. The final grades of those who had completed the writing exercise were lower than those who had not in the first-semester course, and the same in the second-semester course. These results suggest that in the first semester the writing exercise was not beneficial to either gender. In summary, the values intervention was useful for helping women improve their concept inventory scores, exam scores, and grades in physics in the initial study. However, a subsequent study at the same institution showed that this did not improve women's concept inventory scores and a subsequent study at another institution showed that in one course it did not improve concept inventory scores for either gender.

Overall, the work by Miyake et al. [29] and Kost-Smith et al. [30] indicate that stereotype threat is likely a factor contributing to the gender gap in women's exam scores and grades in physics, but it is unclear how this stereotype threat may contribute to the gender gap on concept inventories. The values affirmation writing exercise improves women's performance on concept inventories, exams, and grades in some cases while not in others, although the reason for these differences is not understood. Furthermore, looking for simple relationships between how demographic data were collected and how many women were in the class is not sufficient to observe the effect of stereotype threat on concept inventory scores. This leads us to believe that stereotype threat is a complex phenomenon that cannot be simply described by looking at these factors, but may influence the gender gap on concept inventories in physics.

### F. Construction of test questions

It has been suggested that the questions on these concept inventories may be written in a way that favors men, for example, the use of contexts such as airplanes and cannon balls that are more familiar to men. Several studies have looked for this kind of gender bias on the FCI using very different methodologies and have found remarkably consistent results. McCullough and Meltzer [8,9] rewrote questions on the FCI to address the same content with more feminine and everyday contexts, e.g., changing a question about an airplane dropping a package to an eagle dropping a fish. They found that in a calculus-based introductory physics class the average scores for women on this revised version of the FCI (gender FCI) were similar to those on the original FCI. Further, the gender gap in average scores on pre- and posttests was similar for the original and gender FCI. They compared the performance of men and women on individual questions on the original and gender FCI. They found that women performed better on gender FCI items 14 and 23, while men did worse on gender FCI item 22 and better on item 29. These differences averaged out so that, overall, rewriting the FCI to have more feminine and everyday contexts did not change the gender gap.

There has also been work using differential item functioning (DIF) statistical methods to identify items on the FCI which favor a gender by calculating the probability that individuals with the same ability from different subgroups will answer the same item correctly [20,32]. Dietz et al. [20] looked at pre- and posttest data from an introductory calculus-based physics class over several semesters and found two questions on the pretest (items 6 and 12) and one on the posttest (item 23) which had significant DIF favoring men and two questions on the posttest (items 4 and 9) with significant DIF favoring women. Popp et al. [32] looked for significant DIF on items of the FCI for posttest data from 95 high school physics classes. They found seven items with significant DIF that favored men and seven items that favored women. They found three items (FCI questions 14, 15, and 23) with "substantial DIF." Items 14 and 23 favored men while item 15 favored women. The authors removed these items from their analysis and still found a high effect size for the difference in scores between men and women on the remaining questions. Neither of these studies found a consistent pattern in individual items favoring men (as measured by significant DIF) and concluded that the items on the FCI were not biased in favor of men. Overall, there is no evidence to show that the gender gap is an artifact of the construction of the questions based on DIF methods.

In summary, these studies used very different analysis methods to come to the same conclusion: while there are small biases in a few individual questions that can be modified by rewording them, on average these biases cancel each other out so that the wording of test questions on the FCI does not seem to be a factor contributing to the gender gap. Also, in both studies (and other studies not discussed here) FCI items 14 [13,34,35] and 23 [13,19,34,35] were found to favor men.

### VI. CONCLUSION AND DISCUSSION

Many aspects of the gender gap on concept inventories in physics have been studied, yet the factors that influence the gap and the way in which they influence it are not clear. On mechanics concept inventories, men's pretest scores are almost always about 12% higher than women's. There is almost always a gender gap on the mechanics posttests favoring men (about 13%), but the size of the gender gap varies more than on the pretest. The way the gender gap changes from pre- to posttest is also quite variable and is not directly related to the class being taught with interactive engagement versus traditional teaching methods. Men usually have a higher normalized gain as well, although this difference in normalized gain (0.06) is much less than the difference in normalized gain between classes taught with traditional versus interactive engagement teaching methods (0.25).

Concept inventories in E&M show a different gender gap pattern than those in mechanics. The pretest scores for E&M concept inventories are probably not meaningful because they do not vary across different populations and are always very low, indicating students likely do not have the requisite knowledge about E&M topics to be measured by these tests. That said, there is still an average gender gap of 3.7% on the E&M pretests and 8.5% on the E&M posttests. The gender gap in normalized gains on E&M concept inventories of 0.06 is similar to the gap on mechanics tests.

Table II summarizes the factors that may contribute to the gender gap on physics concept inventories and the findings pertaining to each factor. Background and preparation have been found to have the largest influence on the gender gap on mechanics and E&M concept inventories. We should interpret this finding with caution, as most of the background factors that account for the gender gap are other tests. It may be that gender differences in test anxiety, test taking skills, or the negative effects of stereotype threat influence these test results and therefore influence the background and preparation measures. We cannot determine if the differences in the background and preparation variables (which explain a large portion of the gender gap) are true differences in preparation or artifacts of the testing situation.

TABLE II. Summary of findings on factors that may influence the gender gap on concept inventories in physics.

| Factor | Result |
|---|---|
| Background and preparation | • FMCE or BEMA pretest, combined math score (ACT, SAT, and/or math placement exam), CLASS pretest, and semester class was taken account for a substantial portion of the gender gap (70% of the FMCE posttest gender gap and 62% of BEMA posttest gender gap) at one institution.<br>• The gender gap was eliminated when 20 pairs of men and women were matched on high school background and pretest scores.<br>• However, Brewe *et al.* found that using SAT score as a covariate did not eliminate the gender gap.<br>• Factors which did not contribute to the gender gap include years of high school physics, years of high school calculus, high school GPA, declared major, ethnicity, self-efficacy, and identity. |
| Gender gaps on other measures | Gender gap on concept inventories (12%–13% in mechanics) is much larger than typical gender gaps on physics exams (0%–4.5%) so the phenomenon of women's poor performance on tests in general is insufficient to explain the gender gap on concept inventories. |
| Difference in personal beliefs and answer "scientist" would give | The difference between students' personal beliefs about physics and their understanding of what scientists think is bigger for women (3%) than men (1%) on the FCI pre- and posttest. But the difference between the two is much smaller than the mechanics concept inventory gender gap of 12%–13%, likely accounting for only a small fraction of the gender gap. |
| Teaching method | Courses taught using interactive engagement techniques exhibit both increases and decreases in the gender gap from pre- to posttest. Interactive engagement techniques are beneficial to both genders in all cases, but it is unclear whether this benefit is greater for either gender or which details of implementation are likely to have a differential impact. |
| Stereotype threat | There is currently insufficient evidence that stereotype threat influences the gender gap based on analyses of the factors most likely to contribute to such an influence. However, stereotype threat is a complex phenomenon that cannot be reduced to a single factor, so it is possible that there is an effect that involves more factors than have been studied.<br>• Whether or not students' gender information is collected immediately before the test does not appear to influence the gender gap.<br>• There is insufficient data to determine if the gender gap could be reduced if the instructor or a large majority of students were female. There is no evidence that the fraction of women in a class influences the gender gap for classes with up to 55% women.<br>• A values affirmation writing exercise improves women's performance on concept inventories, exams, and grades in some cases and not others, and the reasons for these differences are not understood. |
| Question wording | A few FCI questions have small gender biases that can be modified by rewording them, but on average these biases cancel each other out so that the wording of test questions does not appear to be a factor contributing to the gender gap. |

Gender gaps on other measures and differences in personal beliefs and the answer a scientist would give have some influence on the gender gap, but do not fully explain it. There have been mixed findings on how teaching methods and stereotype threat influence the gender gap. The context and construction of the test questions does not seem to influence the gender gap.

### A. Implications for instructors

Instructors should be aware of the existence of the gender gap on concept inventories in physics and the complex nature of this phenomenon. There is a large and consistent gender gap on concept inventories in physics, suggesting many subtle and complex sources of bias towards women in our educational system, both in physics classes and beyond. Instructors should recognize that such biases exist and should work to eliminate them, while recognizing that there is no simple guaranteed solution to do so. There is some evidence that a large portion of the gender gap may be explained by gender differences in background and preparation, suggesting a need for better science, technology, engineering, and mathematics preparation for girls early in their schooling. However, it is unclear whether this evidence is generalizable beyond the institution of the original studies or to what degree this result is an artifact of test anxiety or stereotype threat. There is evidence that, on average, women do worse than men on exams and better than men on homework, suggesting that instructors should carefully consider the weighting of these types of assessments in assigning grades.

While promising work has been done indicating methods to reduce or eliminate the gender gap, for example, interactive engagement teaching methods and values affirmation writing exercises, additional studies do not support these findings. Thus, while it is possible that instructional methods designed to address gender bias and support the learning of female students could make a difference, it is still unclear exactly what these methods should look like. One result that holds consistently across studies is that interactive engagement methods improve student learning over traditional methods for students of both genders. We encourage instructors to use interactive engagement methods to improve learning for all students. We also encourage instructors to supplement these methods with techniques explicitly designed to address the gender gap, but to exercise a healthy skepticism towards such methods and not lean too heavily on the results of one study or to deem the gender gap a "solved problem." Instead, instructors should approach the gender gap as a complex phenomenon with many inputs and interactions.

Instructors should recognize that while the gender gap is not fully understood, it does not appear to be due to systematic bias in the wording of the questions on concept inventories and does not invalidate the use of concept inventories to assess the effectiveness of teaching methods, at least when comparing large differences.

### B. Implications for researchers

Researchers interested in studying the gender gap should recognize that, while substantial research on the gender gap has been done, it is not a solved problem and there are many open questions for further research. It is well established that there is a significant gender gap on physics concept inventories that is consistent across institutions. It is not well established what the causes of this gender gap are or what can be done to eliminate it. Because many studies in this area have produced inconsistent results, this is an area where replication studies are especially needed. In particular, studies at multiple institutions with different populations of students and teaching methods are needed to determine the impacts of background and preparation factors, stereotype threat, and different teaching methods.

Researchers studying the impact of different teaching methods using concept inventories should recognize that gender differences in concept inventory scores could impact the results of such studies, but in subtle and not-well-understood ways. The difference in normalized gain for men and women (0.06) is much smaller than the difference in normalized gain between traditional and interactive engagement teaching methods (0.25), but still large enough that it could impact the results of studies comparing different teaching methods or other factors. When making other comparisons of normalized gain for research purposes, the gender gap could influence research results, although it is not clear in what direction. Since average normalized gains are larger for men than for women, it is possible that having more women in a class could reduce the overall normalized gain for the class, thus making a teaching method appear to be less effective than it might appear in a class with a larger proportion of men. On the other hand, it is possible that if a substantial majority of the students in the class were women, this effect might be mitigated or even reversed. Overall, there are large unexplained variations in the gender gap in normalized gain, suggesting that we do not yet understand how the details of implementation of any teaching method impact the gender gap.

## C. Open questions

There have been many studies investigating the gender gap on concept inventories in physics, yet there are still many open questions. We do not know why the gender gap increases from pre- to posttest in some courses and not others. It is not clear if these differences in gender gap result from characteristics of the teacher, e.g., the gender of the instructor or some kind of instructor gender bias. It is not well understood how the level of interactivity of the teaching method influences the gender gap and, if it does, what specific aspects of the method are most important. Another open question is how the dynamics of student interactions and attitudes influences the gender gap. It is also unclear how stereotype threat influences female physics students and how we can mitigate this effect consistently. These questions should be investigated in future studies and the gender gap on concept inventories should not be considered a well-understood or solved problem.